\documentclass[cits]{PoS}
\usepackage{graphicx}

\title{Opacity in parsec-scale jets of active galactic nuclei:
VLBA study from 1.4 to 15 GHz}

\ShortTitle{Opacity in parsec-scale jets of active galactic nuclei}

\author{\speaker{Yuri~Y.~Kovalev}\\
        Max-Planck-Institut f\"ur Radioastronomie, Auf dem H\"ugel 69, 53123 Bonn, Germany\\
        Astro Space Center of Lebedev Physical Institute, Profsoyuznaya 84/32, 117997 Moscow, Russia\\
        E-mail: \email{ykovalev@mpifr.de}}

\author{Alexander~B.~Pushkarev\\
        Max-Planck-Institut f\"ur Radioastronomie, Auf dem H\"ugel 69, 53123 Bonn, Germany \\
        Pulkovo Observatory, Pulkovskoe Chaussee 65/1, 196140 St. Petersburg, Russia\\
        Crimean Astrophysical Observatory, 98409 Nauchny, Crimea, Ukraine\\
        E-mail: \email{apushkar@mpifr.de}}

\author{Andrei~P.~Lobanov\\
        Max-Planck-Institut f\"ur Radioastronomie, Auf dem H\"ugel 69, 53123 Bonn, Germany\\
        E-mail: \email{alobanov@mpifr.de}}

\author{Kirill~V.~Sokolovsky\\
        Max-Planck-Institut f\"ur Radioastronomie, Auf dem H\"ugel 69, 53123 Bonn, Germany\\
        Astro Space Center of Lebedev Physical Institute, Profsoyuznaya 84/32, 117997 Moscow, Russia\\
        E-mail: \email{ksokolov@mpifr.de}}

\abstract{
In extragalactic jets, the apparent position of the bright/narrow end
(the core) depends on the observing frequency, owing to synchrotron
self-absorption and external absorption. The effect must be taken into
account in order to achieve unbiased results from multi-frequency VLBI
data on AGN jets. Multi-frequency core shift measurements supplemented
by other data enable estimating the absolute geometry and a number of
fundamental physical properties of the jets and their environment. We
have previously measured the shift between 13 and 3.6 cm in a sample of
29 AGNs to range between 0 and 1.4 mas. In these proceedings, we present
and discuss first results of our follow-up study using
VLBA between 1.4 and 15.4~GHz.
}

\FullConference{The 9th European VLBI Network Symposium on The role of VLBI in the Golden Age for Radio Astronomy and EVN Users Meeting\\
		 September 23-26, 2008\\
		 Bologna, Italy}

\begin{document}

\section{Introduction}

In VLBI images of relativistic jets, the location of the narrow end of
the jet (branded the ``core'') is fundamentally determined by absorption
in the radio emitting plasma itself (synchrotron self-absorption) and/or
in the material surrounding the flow
\cite{BlandfordKonigl79,Koenigl81,L98} and can be further modified by
strong pressure and density gradients in the flow \cite{L98}. At any
given observing frequency, $\nu$, the core is located in the jet region
with the optical depth $\tau_s(\nu)\approx 1$, which causes its absolute
position, $r_\mathrm{c}$, to shift $\propto \nu^{-1/k_\mathrm{r}}$. If
the core is self-absorbed and in equipartition, $k_\mathrm{r}=1$
\cite{BlandfordKonigl79}; $k_\mathrm{r}$ can be larger in the presence
of external absorption or pressure/density gradients in the flow
\cite{L98}.

Changes of the core position measured between three or more frequencies
can be used for determining the value of $k_\mathrm{r}$, estimating the
strength of the magnetic field in the nuclear region and the offset of
the observed core positions from the true base of the jet \cite{L98}.
The power index $k_\mathrm{r}$ itself can vary with frequency due to
pressure and density gradients or absorption in the surrounding medium,
most likely, associated with the broad-line region.

If the core shifts and $k_\mathrm{r}$ are measured between four, or
more, frequencies, the following can be addressed in detail.
The magnetic field distribution can be reconstructed in the
ultra-compact region of the jet and estimates of the total (kinetic plus
magnetic field) power, the synchrotron luminosity, $L_\mathrm{syn}$,
and the maximum
brightness temperature, $T_\mathrm{b,max}$ in the jet can be made.
In addition, the ratio of particle energy and magnetic field energy can
be estimated from the derived $T_\mathrm{b,max}$. This would
enable testing the K\"onigl model \cite{Koenigl81} and several of its later
modifications (e.g., \cite{HM86,BM96}).
The location of the central engine and the geometry of the jet can be
determined. Estimation of the distance from the nucleus to the jet
origin will enable constraining the self-similar jet model
\cite{Marscher95} and the particle-cascade model \cite{BL95}.

Previously, we have found \cite{Kovalev_cs_2008} that the shift of the
VLBI core position between 2.3 and 8.6~GHz can be as much as 1.4~mas
(median value 0.44~mas) for a sample of 29~AGN. It was also found that
nuclear flares result in temporal variability of the shift. See
\cite{Kovalev_cs_2008} and references therein for more discussion of the
core shift studies as well as very recent papers by \cite{KLP08,OSG09}.
First selected results from our follow~up study of this effect in
luminous extragalactic jets are presented in these proceedings.

\section{Observational data and core shift measurements}

\begin{figure}[p]
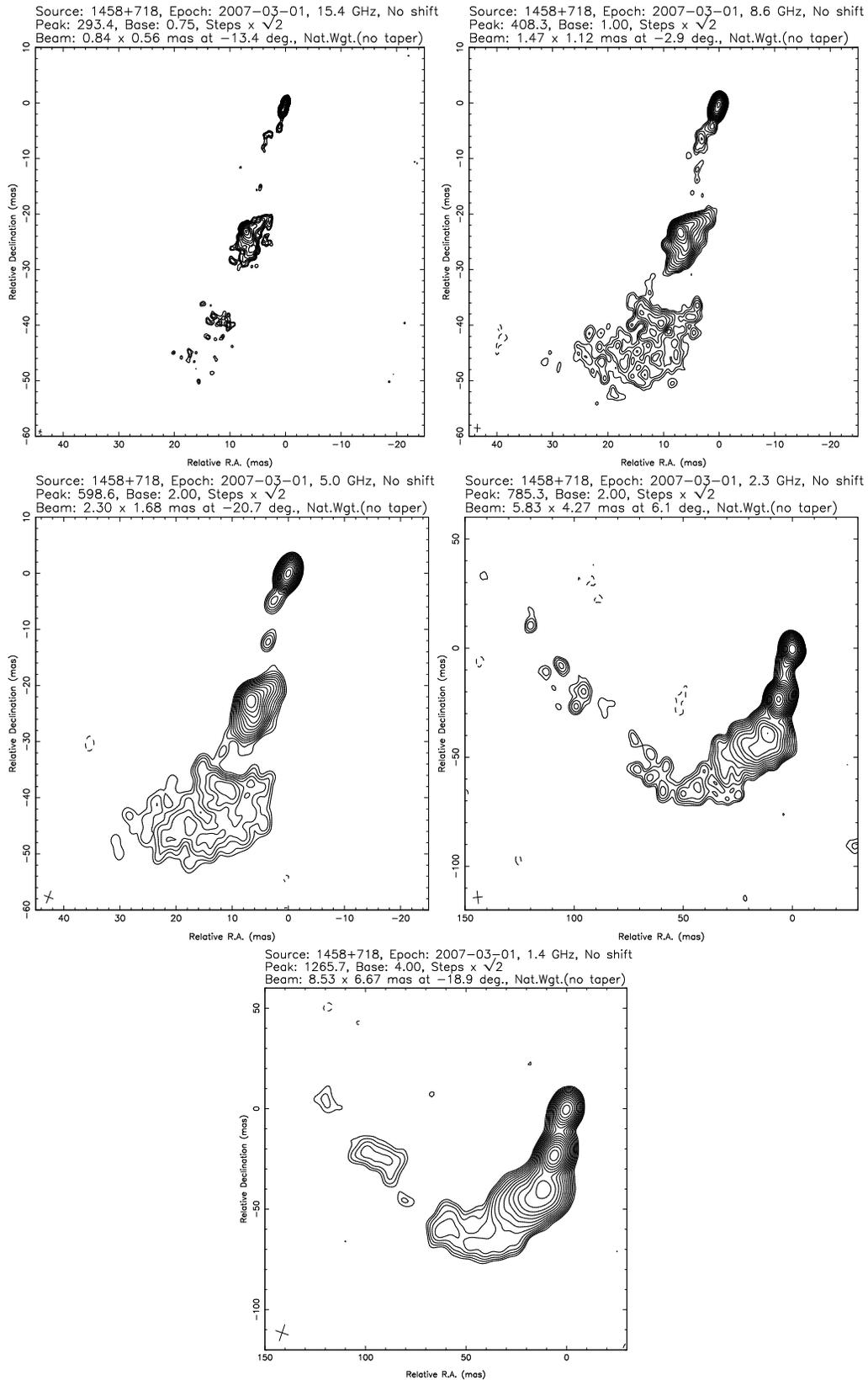

\center
\resizebox{0.9\hsize}{!}{
 \includegraphics[angle=-90,trim=0cm 0cm -0.2cm 0cm,clip=true]{1458+718_U_2007_03_01_pus_map.ps}
 \includegraphics[angle=-90,trim=0cm 0cm -0.2cm 0cm,clip=true]{1458+718_X_2007_03_01_pus_map.ps}
}
\resizebox{0.9\hsize}{!}{
 \includegraphics[angle=-90,trim=0cm 0cm -0.2cm 0cm,clip=true]{1458+718_C_2007_03_01_pus_map.ps}
 \includegraphics[angle=-90,trim=0cm 0cm -0.2cm 0cm,clip=true]{1458+718_S_2007_03_01_pus_map.ps}
}
\includegraphics[width=0.45\textwidth,angle=-90,trim=0cm 0cm 0cm 0cm,clip=true]{1458+718_L_2007_03_01_pus_map.ps}
\caption{\label{f:1458maps}
Stokes I images for frequency bands from 15.4 to 1.4 GHz (top to bottom) 
for the quasar
3C\,309.1. Basic parameters can be found on top of each image.
The lowest contour is plotted on the level `Base' (mJy/beam).
Restoring beam is shown in the left bottom corner of every map.
}
\end{figure}

We have selected the twenty most prominent targets from
\cite{Kovalev_cs_2008} and observed them in a dedicated VLBA experiment
in 2007 (code BK\,134, 96\,hr in total) between 1.4 and 15.4 GHz at nine
separate frequencies with 256~Mbps recording rate. Example Stokes I
images for one of the targets, the compact steep spectrum (CSS) quasar
3C\,309.1 (1458+718) are presented in Figure~\ref{f:1458maps}. 

In order to test even more sources with high resolution and dynamic
range, we also started an analysis of the MOJAVE-2\footnote{See {\tt
http://www.physics.purdue.edu/MOJAVE/}} VLBA observations of 192 radio
loud extragalactic jets, see sample description in \cite{Lister_etal09}.
The MOJAVE-2 observations have
happened in 2006 at four frequencies of 8.1, 8.4, 12.1, \& 15.4~GHz.
Following \cite{Kovalev_cs_2008}, we applied
the self-referencing method to measure the core shift in the data
presented.

\clearpage
\section{Selected Results}
\subsection{1.4-15.4 GHz measurements, quasar 3C\,309.1}

\begin{figure}[b]
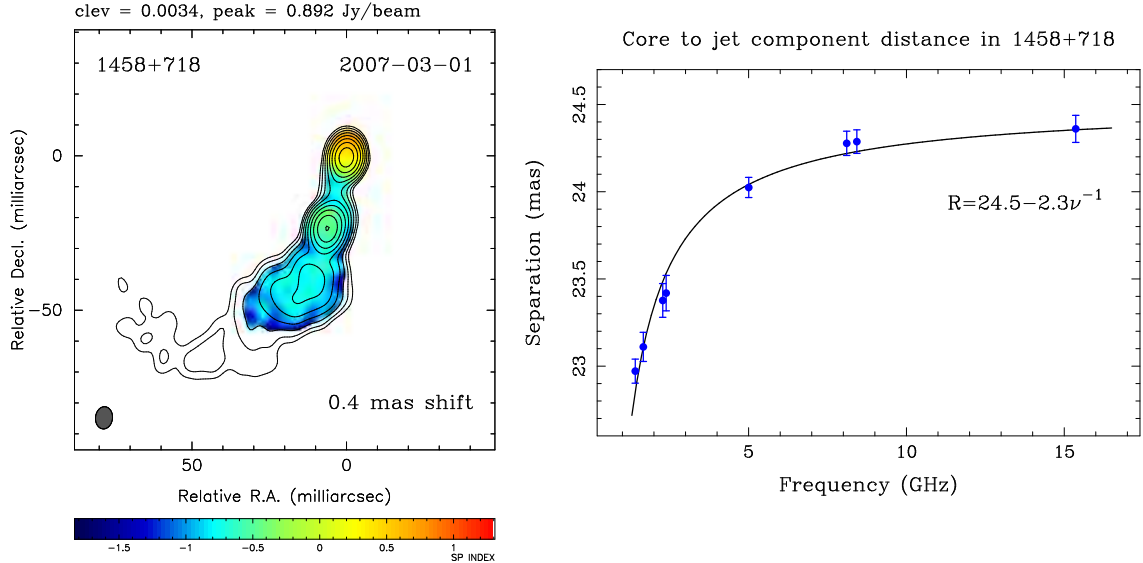

\center
\resizebox{\hsize}{!}{
 \includegraphics[angle=0,trim=0cm 2.5cm 0cm 2cm,clip=true]{1458+718.s2.2007_03_01_spi_l1s2_2.ps}
 \includegraphics[angle=-90,trim=21cm 0cm 0cm 0cm,clip=false]{1458+718_coreshift.ps}
}
\caption{\label{f:1458shift}
Results for the CSS quasar 3C\,309.1.
{\em Left:} Spectral index $\alpha$ ($S\propto\nu^\alpha$) image between 
1.4 and 2.4~GHz, 0.4~mas shift applied. The restoring beam is shown in
the left bottom corner. Spectral index is shown by color while
contours represent the 2.4~GHz CLEAN image.
{\em Right:} Distance between a jet feature centroid position and the
core versus frequency. The curve is fitted to the data for the
pure synchrotron self-absorption case ($k_\mathrm{r}=1$).
}
\end{figure}

Results of the multi-frequency core shift measurements and spectral
index image corrected for the shift are presented in
Figure~\ref{f:1458shift} for 3C\,309.1. This CSS quasar is located at
redshift 0.905 \cite{HB89}. One milliarcsecond corresponds to 7.82~pc.
We adopt the following parameters from \cite{L98}: jet Lorentz factor
$\Gamma_\mathrm{jet}=5$, observing angle of the jet
$\vartheta_\mathrm{jet}=20^\circ$, jet opening angle $\phi=2^\circ$. We
also use the value $\alpha=-0.6$ for the spectral index of the jet.
Applying the model and method by \cite{L98} we study the frequency
dependence of the opacity in the jet and we derive the basic physical
properties of the flow at the location of the VLBI core. The offset
measure $\Omega_{\mathrm r\nu}$~\cite{L98} might be slightly affected by
the blending below 5~GHz for 3C\,309.1. In view of this, the higher
frequency measurements can be employed to derive the physical properties
of the jet. Using the shifts between 5, 8, and 15~GHz and the jet
parameters summarized above, we estimate the distance from the 15~GHz
core to the central super-massive black hole to be $5\pm2$~pc. The
magnetic field at a distance of 1~parsec from the nucleus is estimated
to be $2.3\pm0.5$~G, adopting the electron density
$N_e=17,000$~cm$^{-3}$~\cite{kus93}. This is comparable with the
determined equipartition magnetic field $B_\mathrm{eq}=1.7\pm0.8$~G. The
estimate above corresponds to a magnetic field of $0.1\pm0.2$~G in the 
apparent core region observed at 15~GHz. We also estimate the total
luminosity $L_\mathrm{tot} = (1.1 \pm 0.2) \times 10^{47}$~erg\,s$^{-1}$
and the synchrotron luminosity $L_\mathrm{syn} = (1.4 \pm 0.3) \times
10^{46}$~erg\,s$^{-1}$, for the compact, unresolved section of the jet.
The maximum brightness temperature predicted for this section of the jet
is $(8 \pm 6) \times 10^{11}$~K, which suggest that the jet emission is
dominated by the energy release due to the Compton losses.

\clearpage
\subsection{8.1 to 15.4 GHz measurements, BL~Lacertae object 0716+714}

\begin{figure}[t]
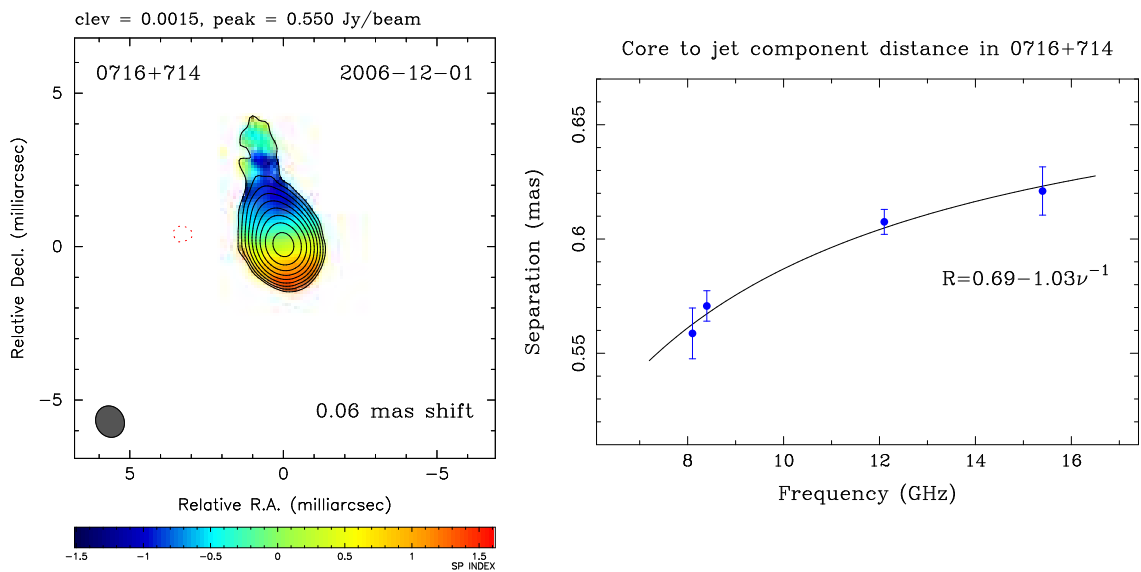

\center
\resizebox{\hsize}{!}{
 \includegraphics[angle=0,trim=0cm 2.5cm 0cm 2cm,clip=true]{0716+714.u.2006_12_01_jets_aligned_spi_xu.ps}
 \includegraphics[angle=-90,trim=21cm 0cm -2cm 0cm,clip=false]{0716_coreshift.ps}
}
\caption{\label{f:0716shift}
Results for the BL~Lac object 0716+714.
{\em Left:} Spectral index $\alpha$ image between 8 and~15 GHz,
0.06 mas shift applied.
The restoring beam is shown in the left bottom corner.
Spectral index is shown by color while
contours represent the 15.4~GHz CLEAN image.
{\em Right:} Distance between a jet feature centroid position and the core
versus frequency, $k_\mathrm{r}=1$ is used for the fitted curve.
}
\end{figure}

Figure~\ref{f:0716shift} reports successful core shift measurements in
the very interesting intra-day variable BL~Lacertae object 0716+714
(e.g., \cite{Wagner_etal96,Rastorgueva_etal09}, and references therein).
This source is located at redshift 0.31 \cite{Nilsson_etal08}, one
milliarcsecond corresponds to 4.52~pc. The core shift between 8.1 and
15.4~GHz is found to be only about 0.06~mas. However, even this
relatively small value affects spectral index imaging significantly if
not accounted for.

\section{Summary}

The core shift effect is important to account for performing multi
frequency VLBI analysis of compact extragalactic jets and using these
objects as reference points in astrometry. It provides a useful tool to
derive and study physical properties of the apparent compact
relativistic jet origin and the absolute jet geometry. In these
proceedings we described our multi-frequency follow~up study of this
effect. We reported selected preliminary results and confirmed previous
findings.

\clearpage
%\medskip
\noindent
{\bf Acknowledgments.} 
This research has made use of data from the MOJAVE database that is
maintained by the MOJAVE team \cite{Lister_etal09}. Part of this
project was done while Y.~Y.~Kovalev was working as a research fellow of
the Alexander von~Humboldt Foundation. K.~V.~Sokolovsky is supported by
the International Max Planck Research School (IMPRS) for
Astronomy and Astrophysics; his participation in the 9th EVN Symposium was
partly supported by funding from the European Community's sixth
Framework Program under RadioNet R113CT~2003~5058187. The VLBA is a
facility of the National Science Foundation operated by the National
Radio Astronomy Observatory under cooperative agreement with Associated
Universities, Inc. This research has made use of NASA's Astrophysics
Data System and NASA/IPAC Extragalactic Database (NED) 
which is operated by the Jet Propulsion Laboratory, California Institute
of Technology, under contract with the National Aeronautics and Space
Administration.

%\bibliographystyle{iopart-num}
%\bibliography{yyk}

\end{document}